# Serializability, not Serial: Concurrency Control and Availability in Multi-Datacenter Datastores

Stacy Patterson[1]  Aaron J. Elmore[2]  Faisal Nawab[2]  Divyakant Agrawal[2]  Amr El Abbadi[2]

[1]Department of Electrical Engineering
Technion - Israel Institute of Technology
Haifa, 32000, Israel
stacyp@ee.technion.ac.il

[2]Department of Computer Science
University of California, Santa Barbara
Santa Barbara, CA 93106
{aelmore,nawab,agrawal,amr}@cs.ucsb.edu

## ABSTRACT

We present a framework for concurrency control and availability in multi-datacenter datastores. While we consider Google's Megastore as our motivating example, we define general abstractions for key components, making our solution extensible to any system that satisfies the abstraction properties. We first develop and analyze a transaction management and replication protocol based on a straightforward implementation of the Paxos algorithm. Our investigation reveals that this protocol acts as a concurrency prevention mechanism rather than a concurrency control mechanism. We then propose an enhanced protocol called Paxos with Combination and Promotion (Paxos-CP) that provides true transaction concurrency while requiring the same per instance message complexity as the basic Paxos protocol. Finally, we compare the performance of Paxos and Paxos-CP in a multi-datacenter experimental study, and we demonstrate that Paxos-CP results in significantly fewer aborted transactions than basic Paxos.

## 1. INTRODUCTION

Cloud computing has the potential to become the foundation for most information technology architectures. It offers application developers access to seemingly infinite storage, compute, and network resources, all on a pay-per-use basis. While the appeal of the cloud computing model is obvious from a financial perspective, its success also depends on the ability of clouds to provide reliable, scalable services that support the features developers need. In particular, it is important that cloud datastores, such as Google's BigTable [8] and Amazon's SimpleDB [1], provide support for various types of data consistency and guarantee the availability of application data in the face of failures.

Initially, cloud datastores provided only *eventually consistent* update operations guaranteeing that updates would eventually propagate to all replicas. While these datastores were highly scalable, developers found it difficult to create applications within the eventual consistency model [20]. Many cloud providers then introduced support for atomic access to individual data items, in essence, providing *strong consistency* guarantees. This consistency level has become a standard feature that is offered in most cloud datastore implementations, including BigTable, SimpleDB, and Apache HBase [16]. Strong consistency of single data items is sufficient for many applications. However, if several data items must be updated atomically, the burden to implement this atomic action in a scalable, fault tolerant manner lies with the software developer. Several recent works have addressed the problem of implementing ACID transactions in cloud datastores [2, 10, 11], and, while full transaction support remains a scalability challenge, these works demonstrate that transactions are feasible so long as the number of tuples that are transactionally related is not "too big".

While many solutions have been developed to provide consistency and fault tolerance in cloud datastores that are hosted within a single data center, these solutions are of no help if the entire datacenter becomes unavailable. For example, in April 2011, a software error brought down one of Amazon's EC2 availability zones and caused service disruption in the U.S. East Region [24]. As a result, major web sites like Reddit, Foursquare, and Quora were unavailable for hours to days [5]. And, in August 2011, lightning caused Microsoft and Amazon clouds in Dublin [15] to go offline for hours. In both instances, there were errors in the recovery process, and it was not possible to restore a consistent snapshot of some application data.

These recent outages demonstrate the need for replication of application data at multiple datacenters as well as the importance of using provably correct protocols for performing this replication. In a recent work, Baker et al. describe Megastore, Google's approach to providing transactions in the cloud with full replication at multiple datacenters [2]. Megastore is implemented on top of BigTable and provides support for ACID transactions over small sets of data items called *entity groups*. It uses multi-version concurrency control and a replicated write-ahead log. Replication is performed using the Paxos algorithm [18] to ensure consistency even with unreliable communication and datacenter outages. While the paper presents an overview of the Megastore system, it lacks the formality and detail required to verify Megastore's correctness. We assert that such analysis is needed for systems like Megastore, especially in light of the outages described above and the widely acknowledged difficulties associated with understanding and implementing the Paxos algorithm [7, 19, 25].

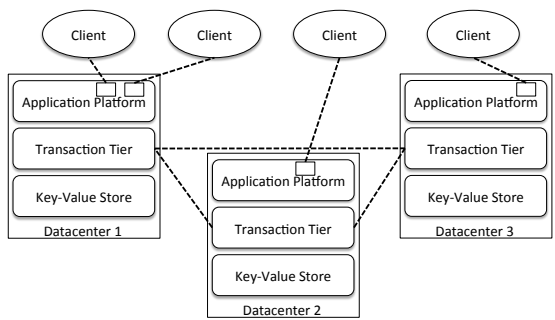

**Figure 1: System architecture for transactional cloud datastore with full replication at multiple datacenters.**

In this work, we address the need for formal analysis of replication and concurrency control in transactional cloud datastores. We define and analyze several Paxos-based protocols for replication and transaction management in the multi-datacenter setting. While we take Megastore as our motivating example, we define general abstractions for each of the key components, and we use these abstractions in our protocol design and analysis. The specific contributions of our work are:

- We provide a formal description of the Paxos protocol for replication and concurrency control, and we prove its correctness. Through our analysis, we also show that the Paxos protocol, as implemented in Megastore, aborts transactions that could be safely committed. In essence, it acts as a concurrency prevention mechanism rather than a concurrency control mechanism.

- We propose an enhanced replication and concurrency control protocol that we call *Paxos with Combination and Promotion* (Paxos-CP). Paxos-CP enables true transaction concurrency, with the same per-instance message complexity as the original Paxos protocol.

- We compare the performance of Paxos and Paxos-CP in a multi-datacenter experimental study, and we demonstrate the benefits of our enhanced Paxos protocol.

The remainder of this paper is organized as follows. In Section 2, we give an overview of the design of the cloud datastore including the data model and reference architecture. Section 3 summarizes the theoretical foundations that we use to analyze the correctness of the transactional cloud datastore. In Section 4, we present the details of the transaction manager, including the basic Paxos commit protocol, and we prove its correctness. In Section 5, we present our extended Paxos commit protocol that allows for transaction concurrency, and we prove the correctness of this protocol. We present evaluation results comparing the basic and extended Paxos commit protocols in Section 6. In Section 7, we discuss related work, and we conclude in Section 8.

## 2. SYSTEM OVERVIEW

We consider a cloud platform that consists of a small number of datacenters, as shown in Figure 1. Each application is replicated in the application platform of each datacenter. The application uses the transactional datastore to store all non-volatile application state, and this data is replicated at every datacenter. Therefore, clients can access any copy of the application at any datacenter, and the system should appear as if there is only one copy of the application. Multiple clients can access the same application at the same time and can use any communication method supported by the application (e.g. TCP or UDP). We first describe the data model for application data and metadata, and then, we give an overview of the architecture within each datacenter.

### 2.1 Data Model

Application data is stored in a datastore that has a *key-value store* as its foundation. Each data item consists of a unique *key* and its corresponding *value*, an arbitrary number of *attributes* (also called columns). An application specifies which data items are transactionally related, i.e., which data items can be accessed within a single transaction. A set of data items that can be accessed within a single transaction is called a *transaction group* and is identified by a *transaction group key* that is unique within and across applications. Each application can have multiple transaction groups, and the assignment of data items to transaction groups is performed *a priori*. For clarity of development, we assume that every data item belongs to exactly one transaction group and can only be accessed within the scope of a transaction. A client may execute transactions on multiple transaction groups concurrently, but the system does not support multi-transaction group transactions; each transaction succeeds or fails independent of the others, and there is no guarantee of global serializability across transaction groups.

As discussed in the introduction, it is possible to implement ACID transactions in a cloud-scale database provided the transactional group is not too big. What size actually qualifies as too big depends on the datastore implementation and physical architecture as well as the level of contention for items in the same transaction group. We explore the relationship between performance and level of contention in Section 6.

### 2.2 Datacenter Architecture

We follow Megastore's design whereby every datacenter is able to process transactions as long as it is alive. This is in contrast to a master-slave approach where a single master datacenter handles all transactions. As shown in Figure 1, each datacenter is divided into three logical tiers, a *key-value store*, a *transaction tier*, and an *application platform*. Many existing cloud services already provide a key-value store and an application platform. We do not tie our system to any single implementation; we only require that these tiers meet the requirements outlined in the descriptions below.

**Key-value store.** At the foundation is a key-value store. Physically, the key-value store is implemented by a large collection of servers. However, the key-value store also provides a naming service so that clients, in this case the transaction tier, can access the key-value store as a single logical entity. Many key-value store implementations exist, each with its own approach to consistency, elasticity, and fault-tolerance. We do not tie our approach to any one implementation; we only require that the key-value store provides atomic access to individual rows and stores of multiple versions of each row. Specifically, we assume that the key-value store provides the following operations, each of which is executed atomically.

- **read(in:** *key*, *timestamp*; **out** *value*): Read the value for the row for the specified *key*. The read operation returns the most recent version of the row with a timestamp less that or equal to *timestamp*. If no timestamp is specified, the read operation returns the most recent version.

- **write(in:** *key*, *value*, *timestamp*): Write the value to the row specified by *key* by creating a new version of the row with a timestamp specified by *timestamp*. If a version with greater timestamp exists, an error is returned. If no timestamp is specified, a timestamp is generated that is greater than the timestamp for any existing version.

- **checkAndWrite(in:** *key.testAttribute*, *testValue*, *key*, *value*; **out:** *status*): If the latest version of the row has the attribute *testAttribute* equal to the value *testValue*, then it performs **write**(*key*, *value*) operation and returns with success status. Otherwise, the operation is not performed and an error is returned.

We note that these features are supported by many well-known key-value store implementations, including BigTable [8], HBase [16], and Amazon SimpleDB [1]. In each of these operations, the *timestamp* parameter is a logical timestamp. Its value is specified by the caller (in our case, the transaction tier) as part of the concurrency control and replication protocol. In Section 3.2, we explain how the timestamp is determined.

**Transaction tier.** The second tier is the transaction tier. Every transaction tier can communicate with every other transaction tier, though communication may be unreliable. If a message is sent from one transaction tier to another, either the message arrives before a known timeout or it is lost. Individual transaction tiers may go offline and come back online without notice. The transaction tier is responsible for ensuring a serializable execution for concurrent transactions both within a single datacenter and across multiple datacenters. This tier is also responsible for replicating transactions to all datacenters.

The transaction tier implements an *optimistic concurrency control* protocol. For each transaction, read operations are executed on the datastore and write operations are performed on a local copy. Only on transaction commit are the write operations stored in the datastore. We assume that each application instance has at most one active transaction per transaction group.

While the transaction tier is logically a single tier, it is implemented as two entities, a *Transaction Client* that is used by the application instance and a *Transaction Service* that handles requests from the Transaction Client. The Transaction Client provides the following standard transaction API for use by applications in the application platform.

- **begin(in:** *groupKey*): Start a new transaction on the transaction group identified by *groupKey*.

- **read(in:** *groupKey*, *key*; **out:** *value*): Read the value of *key* from the datastore.

- **write(in:** *groupKey, key, value*): Write (*key, value*) to the transactional group specified by *groupKey* in the datastore.

- **commit(in:** *groupKey*; **out:** *status*): Try to commit the transaction. Returns the status of the commit operation, either *commit* or *abort*.

The Transaction Client communicates with the Transaction Service to implement the **begin** and **read** operations. If the transaction is read-only, **commit** automatically succeeds, and no communication with the Transaction Service is needed. If the transaction contains write operations, the Transaction Client and Transaction Services communicate to determine if the transaction can be committed and perform the commit and replication protocol.

The Transaction Service handles each client request in its own service process, and these processes are stateless. If a Transaction Client cannot access the Transaction Service within its own datacenter, it can access the Transaction Service in another datacenter to handle a request. Since the Transaction Service is stateless, the number of instances can be increased as needed to satisfy client requests. However, there is a trade-off between the number of concurrent transactions and the number of transactions that will be aborted by the concurrency control mechanism. We explore this trade-off through simulations in Section 6.

**Application platform.** Applications are hosted within the application platform, and each application is replicated at every datacenter. When a client executes an application, it runs its own copy of the application in its own thread, e.g., the application platform acts as a multi-threaded server that spawns a new thread to handle each client request. A transaction is executed in a single application instance on a single datacenter, and the state for active (uncommitted) transactions exists only within the scope of this instance. If an application platform becomes unavailable, its active transactions are implicitly aborted, and their volatile state is lost. Any data that must be available across multiple client requests must be stored in the transactional datastore via committed transactions.

In the next section, we give the theoretical background necessary for analyzing the transaction tier implementation, and we define the properties that must be satisfied for an implementation to be provably correct.

## 3. THEORETICAL FOUNDATIONS

In our transactional datastore, each datacenter has its own multi-version key-value store. Every data item is replicated at the key-value store within each datacenter, and so, there are both multiple copies and multiple versions of each data item. Yet, when a client (an application instance) executes a transaction, it should appear that (1) there is only one copy and one version of each data item, and (2) within the scope of its transaction, the client is the only one accessing those data items. These two properties are captured by the notion of *one-copy serializability* [4]. We implement one-copy serializability in a multi-datacenter setting using a fully replicated write-ahead log. We briefly formalize the concepts of one-copy serializability and write-ahead logging below.

### 3.1 One-Copy Serializability

In a single copy, single version (SCSV) datastore, a *transaction* is a partially ordered set of read and write operations, terminated by either a single *commit* or single *abort* operation. An *SCSV history* over a set of transactions is the union of the operations of the transactions along with a partial order. This partial order maintains the order of operations within each transaction and specifies an order for all conflicting operations (two operations conflict if they operate on the same data item and at least one of them is a write). We say that a transaction $t$ *reads-x-from* transaction $s$ if $s$ writes $x$ before $t$ reads $x$, and no other transaction writes $x$ in between those two operations.

| α | β | γ | δ |   |
|---|---|---|---|---|
| 1 | 2 | 3 | 4 | 5 |

**Figure 2: Log for a single transaction group at a single datacenter. The last committed transaction is written in position 4, the *read position*. The next available position is 5, the *commit position*.**

In a multi-version, multi-copy (MVMC) datastore, when a client performs a read operation, it reads a single version of a single copy of a data item, When a write operation is applied to the cloud datastore, a new version of the item is created at one or more datacenters. An *MVMC transaction* is a partially ordered set of read and write operations, with their corresponding version and copy attributes, ending with a single *commit* or a single *abort* operation. We say a transaction $t$ *reads-x-from* transaction $s$ if $t$ reads the version of $x$ (at one copy) that was written by $s$ (at one or more copies). An *MVMC history* is a set of MVMC transactions with a partial order. The partial order obeys the order of operations within each transaction and maintains the *reads-from* relation, i.e., if transaction $t$ reads version $i$ of $x$ from transaction $s$ at copy $A$, then the write of version $i$ at copy $A$ precedes the read of version $i$ at copy $A$, and no other write occurs between these operations at copy $A$.

**Definition 1** *A multi-version, multi-copy history H is **one-copy serializable** if there exists a single copy, single version serial history S such that*

1. *H and S have the same operations.*
2. *$t_i$ reads-x-from $t_j$ in H iff $t_i$ reads-x-from $t_j$ in S.*

Our goal is to prove that the system and protocols defined in this paper guarantee one-copy serializability in a multi-version, multi-copy datastore.

### 3.2 A Replicated Write-Ahead Log

Our system implements an optimistic concurrency control protocol with a write-ahead log. In addition to its set of data items, each transaction group has its own write-ahead log that is replicated at all datacenters. The write ahead log is divided into *log positions* which are uniquely numbered in increasing order, as shown in Figure 2. When a transaction that contains write operations commits, its operations are written into a single log position, the *commit position*. Read-only transactions are not recorded in the log. For each write in the committed transaction, the commit log position serves as the timestamp for the corresponding write operation in the key-value store. While the log is updated at commit time, these write operations may be performed later by a background process or as needed to serve a read request.

To guarantee correct execution of transactions, we must be sure that transactions are only written to the log if they are correct with respect to the one-copy serializability property. Formally, we require that our concurrency control protocol maintain the following properties.

**(L1)** The log only contains operations from committed transactions.

**(L2)** For every committed transaction that contain a write operation, all of its operations are contained in a single log position.

**(L3)** An entry will only be created in a log position if the union of this log entry and the complete prefix of the log prior to this log entry is a one-copy serializable history.

We require that transactions are consistently replicated across multiple datacenters. To achieve consistent replication, when a transition commits, we replicate the new log entry at every datacenter. The replication algorithm must satisfy the following property.

**(R1)** No two logs have different values for the same log position.

### 3.3 Transaction Management Correctness

To guarantee correctness of our system, we need two additional assumptions that relate to the handling of read requests.

**(A1)** If an application writes a data item, and then subsequently reads that data item in the same transaction, it will read the value it wrote in the transaction.

**(A2)** Within a transaction, all read operations for data items that were not previously written in that transaction read from the same log position; i.e., the transaction reads the latest writes performed up through the specified read position in the log.

We note that property (A1) is stated for convenience only; (A1) is subsumed by property (L3), since violating (A1) would violate one-copy serializability.

The following theorem shows that the properties defined above are sufficient to guarantee one-copy serializability. We provide a sketch of the proof here. The full proof can be found in a technical report [21].

**Theorem 1** *For the transactional data store with replication at multiple datacenters, if the Transaction Tier guarantees properties (L1) - (L3),(R1), and (A1) - (A2), then the datastore guarantees one-copy serializability.*

PROOF SKETCH. Let $H$ be a history of transactions, and let $k$ be the maximum log position, over all log replicas, that contains a committed transaction from $H$. We define the history $S$ to be the sequence of operations in log positions $1, \ldots, k$ in order. By properties (R1) and (L1) - (L3), $S$ is one-copy serializable history that contains all read/write transactions in $H$ (and respects the reads-from relations of those transactions). Let $T$ be the serial history that is one-copy equivalent to $S$, i.e., $t_i$ reads-x-from $t_j$ in $S$ iff $t_i$-reads-x from $t_j$ in $T$. We form a serial history $U$ that is one-copy equivalent to $H$ by adding the read-only transactions to $T$ as follows. Let $j$ be the log position that was used for the remote read operations, as specified by (A2), and let $t_j$ be the last transaction in $T$ that was written in log position $j$. We form $U$ from $T$ by inserting each read only transaction $t_i$ immediately after its corresponding $t_j$. For multiple read-only transactions that read from the same log position $j$, they can be inserted in any order immediately after $t_j$. □

In the next section, we present the details of the transaction tier implementation and prove that it guarantees the properties defined above.

## 4. THE TRANSACTION TIER

All aspects of transaction management are handled by the two layers of the transaction tier, the Transaction Client and the Transaction Service. Using a transaction library with a standard API, every application instance can act as a Transaction Client. The Transaction Client stores the *readSet* and *writeSet* for each active transaction. Every transaction group has an associated Transaction Service in every datacenter. The Transaction Service provides the interface to the items in the key-value store in each datacenter.

To execute a transaction, the Transaction Client communicates with the Transaction Service in each datacenter to perform the following transaction protocol.

1. When the application invokes **begin**, the Transaction Client contacts the Transaction Service in its local datacenter to determine the position of the last written log entry. We call this the *read position*. The read position indicates the timestamp to use for read operations. If the local Transaction Service is not available, the library contacts Transaction Services in other datacenters until a response is received.

2. When an application invokes a **read** operation, the Transaction Client checks if the same key was written previously within the transaction. If so, the last written value is returned to the application. Otherwise, the Transaction Client sends the read request, along with the read position, to the Transaction Service. For now, we assume that the write-ahead log is up-to-date with respect to the read position. Later, we show how this assumption can be removed. If the log entries up through read position have not yet been applied to the datastore, the Transaction Service applies these operations. The Transaction Service then returns the requested value. If a Transaction Service becomes unavailable, the Transaction Client sends the read request to a Transaction Service in another datacenter. Before returning the value to the application, the Transaction Client adds the corresponding key to *readSet*.

3. **write** operations are handled locally by the Transaction Client by adding (*key*,*value*) to *writeSet*.

4. When the application invokes the **commit** operation, the *readSet* and *writeSet* are combined to form the transaction log entry. The Transaction Client communicates with all available Transaction Services to perform the commit protocol. If the commit is successful, the transaction will be written in the write-ahead logs. After the protocol is complete, the Transaction Client returns COMMIT or ABORT to the application instance.

This implementation of read operations ensures that properties (A1) and (A2) are satisfied. What remains is to define a commit protocol that satisfies the log and replication properties, (L1) - (L3) and (R1). Megastore uses a commit protocol based on the Paxos algorithm, where a single instance of the algorithm is used for each log entry. In Section 4.1, we give a formal description of such a commit protocol, which we call the basic Paxos commit protocol, and in Section 4.2, we prove that this protocol guarantees one copy serializability.

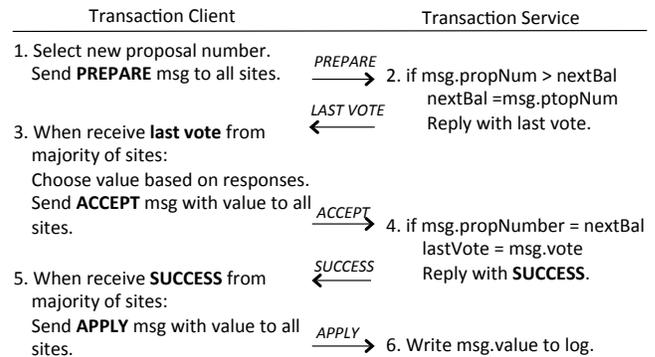

**Figure 3: A single instance of the basic Paxos commit protocol, as executed by the Transaction Client and the Transaction Service.**

### 4.1 The Basic Paxos Commit Protocol

The Paxos algorithm [18] was proposed as a solution for state machine replication in a distributed system with intermittent node failures and message loss. Replicas communicate with each other to reach consensus on a sequence of values, and every site agrees on the same value for every position in the sequence. While the Paxos algorithm can be used to replicate a write-ahead log consisting of a sequence of log entries, it cannot ensure that the log is a one-copy serializable history. Therefore, to use Paxos as a commit protocol, it is necessary to integrate a concurrency control mechanism alongside the replication.

Megastore employs a single instance of the Paxos algorithm[1] to act as both a concurrency control and replication mechanism for each log position. As stated in our transaction protocol, when a transaction begins, the Transaction Client identifies the read position to be used for all read operations. When the Transaction Client attempts to commit the transaction, it tries to commit to a specific log position, the *commit position*, where commit position = read position + 1. All transactions that have the same read position will try to commit to the same commit position. The Paxos commit protocol is used to determine which transaction "wins" the log entry at the commit position and also to replicate the winning transaction to the write-ahead log at every datacenter. The winning transaction receives a COMMIT response to its **commit** request. All other competing transactions receive an ABORT response.

A summary of the steps of the basic Paxos commit protocol is shown in Figure 3, and the pseudocode for the Transaction Service and Transaction Client are given in Algorithms 1 and 2, respectively. One instance of this protocol is executed for every log position. The Transaction Client is responsible for running the protocol by exchanging messages with the Transaction Service in each datacenter. Each Transaction Service stores its state for each Paxos instance in its local key-value store, and it updates this state when it receives messages from Transaction Clients according to the rules defined by the protocol.

---

[1]This single instance of Paxos is more correctly called the Synod Algorithm.

**Algorithm 1:** Steps of Paxos commit protocol implemented by Transaction Service for commit to log position $P$.

```
 1  datastore state for log position P
 2      ⟨nextBal, ballotNumber, value⟩, initially ⟨−1, −1, ⊥⟩

 3  on receive(cid, PREPARE, propNum)
 4      keepTrying ← TRUE
 5      while keepTrying do
 6          (vNextBal, vBallotNumber, vValue) ← kvstore.read(P)
 7          if propNum > vNextBal then
 8              // Only update nextBal in datastore if it has not changed since read.
 9              status ← kvstore.checkAndWrite(P.nextBal, propNum, P.nextBal, vNextBal)
10              if status = SUCCESS then
11                  send(cid, status, vBallotNumber, vValue)
12                  keepTrying ← FALSE
13          else
14              send(cid, FAILURE, vBallotNumber)
15              keepTrying ← FALSE

16  on receive(cid, ACCEPT, propNum, value)
17      // Only write value to datastore if propNum corresponds to most recent update to nextBal for a PREPARE message.
18      status ← kvstore.checkAndWrite(P.⟨ballotNumber, value⟩, ⟨proposalNumber, value⟩, P.nextBal, propNum)
19      send(cid, status)

20  on receive(cid, APPLY, propNum, value)
21      kvstore.write(P.⟨ballotNumber, value⟩, ⟨propNum, value⟩)
```

**Algorithm 2:** Steps of Paxos commit protocol implemented by Transaction Client on commit of value $val$ to log position $P$.

```
22  state
23      propNum, initially 0

24  // PREPARE phase
25  keepTrying ← TRUE
26  propVal ← cval
27  while keepTrying do
28      responseSet ← ∅
29      ackCount ← 0
30      // Loop iterations may be executed in parallel.
31      for each datacenter d do
32          send(d, PREPARE, propNum)
33          while no timeout do
34              (status, num, value) ← receive(i, status, num, val)
                 reponseSet ← responseSet ∪ (status, num, val)
35              if status = SUCCESS then
36                  ackCount ← ackCount + 1
37      if ackCount > (D/2) then
38          keepTrying ← FALSE
39      else
40          sleep for random time period
41          propNum ← nextPropNumber(responseSet, propNum)

42  // ACCEPT phase
43  propValue ← findWinningVal(responseSet, propValue)
44  ackCount ← 0
45  responseSet ← ∅
46
47  Loop iterations may be executed in parallel.
48  for each datacenter d do
49      send(d, ACCEPT, proposalNumber, proposalValue)
50      while no timeout do
51          receive(id, status)
52          if status = 'success' then
53              ackCount ← ackCount + 1
54  if ackCount < majority then
55      sleep for random time period
56      propNum ← nextPropNumber(responseSet, propNum)
57      go to PREPARE phase

58  // APPLY phase
59  //Loop iterations may be executed in parallel.
60  for each datacenter d do
61      send(d, proposalNumber, propValue)
62  if val contained in propVal then
63      return COMMIT
64  else
65      return ABORT

66  function findWinningVal(responseSet, propVal)
67      maxProp ← −1
68      winningValue ← null
69      for (status, num, val) in responseSet do
70          if (num > maxProp) and val ≠ ⊥ then
71              maxProp ← num
72              winningValue ← val
73      if winningValue = ⊥ then
74          winningValue = propVal
75      return winningValue

76  function enhancedFindWinningVal(responseSet, propVal)
77      maxVal ← value in reponseSet with max. num. votes
78      maxVotes ← number of votes for maxVal
79      if maxVotes + (D − |responseSet|) ≤ D/2 then
80          // No winning value so combine.
81          generateCombinedValue(responseSet, propVal)
82      else if (maxVotes > D/2) and
            (propVal not contained in maxVal) then
83          // Another value has already won.
84          try to promote
85      else
86          // Revert to Basic Paxos function.
87          return findWinningVal(responseSet, propVal)
```

On a high level, in the Paxos commit protocol, concurrent transactions compete to get votes from the replicas. The transaction that receives a majority of votes is written to the commit position in the write-ahead log. The first phase of the protocol is the PREPARE phase. The Transaction Client execution of this phase is given in lines 24-41 of Algorithm 2, and the Transaction Service execution is given in lines 3-15 of Algorithm 1. When a Transaction Client wants to commit a transaction, it first selects a proposal number for its commit request. The proposal number must be unique and should be larger than any previously seen proposal number. The client then sends a PREPARE message with this number to all Transaction Services (Step 1). When a Transaction Service receives a PREPARE message, it checks its local key-value to see if it has answered any previous PREPARE message with a larger proposal number. If it has not, the Transaction Service responds to the current PREPARE request by sending the last vote that it cast for the commit position, i.e. the value that it voted should be written to the log (Step 2). If the Transaction Service has not yet cast a vote for the commit position, it sends a response with a null vote. If the Transaction Client receives responses from a majority of Transaction Services within the timeout period, it can proceed to the ACCEPT phase. Otherwise, the Transaction Client must try its PREPARE phase again with a larger proposal number.

In the ACCEPT phase, the Transaction Client proposes a value for the commit log position (lines 43-57 of Algorithm 2). The client first examines the vote information it received from the Transaction Services to determine the proposed value. The Transaction Client must select the value that has the largest proposal number; only if all responses have null values can the client select its own value (see [18]). This determination is handled in the `findWinningVal` function (lines 66-75 of Algorithm 2). The Transaction Client then sends the winning value to all replicas in an ACCEPT message along with its own proposal number (Step 3). When a Transaction Service receives an ACCEPT message (Step 4), it checks if the proposal number is the same as the one to which it responded with its most recent *last vote* message. If so, the Transaction Service casts (or changes) its vote for the value in the message and sends a response to the Transaction Client. Otherwise, the Transaction Service ignores the message. This is shown in lines 16-19 of Algorithm 1.

The Transaction Client collects responses to its ACCEPT messages. If it receives a SUCCESS response from a majority of Transaction Services before the timeout, it has "won" the commit position (lines 50-57 in Algorithm 2). The client then sends the winning value to every Transaction Service in an APPLY message (Step 5 and lines 58-61 in Algorithm 2). If the Transaction Client does not receive enough responses before the timeout, it must begin the protocol again from the PREPARE phase with a larger proposal number. When a Transaction Service receives an APPLY message, it writes the value in that message to the commit position in the write-ahead log (Step 6 and lines 20-21 in Algorithm 1).

We note that when a Transaction Client wins the Paxos instance in Step 5, this does not mean that the client's proposed value will be written to the log. It means that some proposed value, possibly from another transaction, will be written to the log position. Each Transaction Client must execute all steps of the protocol to learn the winning value. The Transaction Client then checks whether the winning value is its own transaction, and if so, it returns a COMMIT status to the application. Otherwise, it returns an ABORT status.

A client will be able to execute the Paxos commit protocol to completion so long as a majority of the Transaction Services are alive, there are only a finite number of proposals for the log position.

**Paxos Optimizations.** A single instance of the Paxos algorithm takes five rounds of messages. In state machine replication, the number of rounds can be reduced to three by designating a *master replica*, a unique leader that remains in charge until it fails (see [18, 7]). In this case, clients skip the PREPARE phase and simply send proposed values to the leader. The leader decides which values are accepted and in what order. With a single leader, communication overhead can be further reduced by clients piggybacking new proposals on acknowledgements of ACCEPT messages. In the context of our system, having a master replica would translate to designating a single site to act as a centralized transaction manager.

Megastore does not use a master replica, but instead designates one leader per log position (see Section 4.4.2 of [2]). We employ the same approach in our system. The leader for a log position is the site local to the application instance that won the previous log position. Before executing the commit protocol, the Transaction Client checks with the leader to see if any other clients have begun the commit protocol for the log position. If the Transaction Client is first, it can bypass the PREPARE phase and begin the protocol at Step 3 with its own value as the proposed value. If the Transaction Client is not first, it must begin the protocol at Step 1. This optimization reduces the number of message rounds to three in cases where there is no contention for the log position. For clarity, we do not include this optimization in the pseudocode in Algorithms 1 and 2. However, we include the optimization in the prototype used in our evaluations.

**Fault Tolerance and Recovery.** If a Transaction Service does not receive all Paxos messages for a log position, it may not know the value for that log position when it receives a read request. If this happens, the Transaction Service executes a Paxos instance for the missing log entry to learn the winning value. Similarly, when the Transaction Service recovers from a failure, it runs Paxos instances to learn the values of log entries for transactions that committed during its outage. If a Transaction Client fails in the middle of the commit protocol, its transaction may be committed or aborted.

### 4.2 Transaction Tier Correctness

We now prove that the transaction tier described in this section implements one-copy serializability.

**Theorem 2** *The transactional datastore with replication at multiple datacenters, implemented using the transaction protocol and basic Paxos commit protocol defined above, guarantees one-copy serializability.*

PROOF. We prove the correctness of the implementation by showing that it guarantees the necessary properties defined in Section 3. With respect to the correctness of read operations, it is clear that the implementation of the read operations stated in the transaction protocol satisfies properties (A1) - (A2). In addition, it has been proven that Paxos algorithm satisfies the log replication property (R1) (see [18]).

By definition of the commit protocol, a transaction commit is equivalent to the transaction's proposed value winning the Paxos instance, and thus, subsequently being written in the write-ahead log. Therefore (L1) is guaranteed. In addition, Paxos guarantees a single value, in its entirety, is written to each log position. In the Paxos commit protocol, the value contains all the operations from a single transaction. Therefore, if a transaction is committed, all of its operations are contained in the single log position designated by the commit position. Thus, (L2) is also guaranteed.

What remains to be shown is that a transaction will only be committed, and therefore, will only be added to the write-ahead log if it does not violate one-copy serializability (property (L3)). Let $H$ be the one-copy serializable history corresponding to the transactions in the write-ahead log up through log position $k$. By definition of the Paxos commit protocol, only one transaction is written in each log position. Therefore, we can trivially define the serial history $\overline{H}$ that is one-copy equivalent to $H$ to be the set of transactions in $H$, ordered by their log positions. Let $t$ be the committed (read/write) transaction with commit position $k+1$. By the transaction protocol, if $t$ reads item $x$, it reads the most recent version of $x$ written in or before log position $k$. Therefore, the history resulting from the commit of transaction $t$ is one-copy equivalent to the serial history where $t$ is appended to the $\overline{H}$. □

While the implementation described in this section is correct, this Paxos commit protocol provides course-grained concurrency control. If two transactions try to commit to the same log position, one will be aborted, regardless of whether the two transactions access the same data items. In some sense, the basic Paxos commit protocol acts as a concurrency prevention mechanism rather than a concurrency control mechanism. In the next section, we show how to extend the protocol to support concurrent read/write transactions, and we prove that this extended protocol also guarantees one-copy serializability.

## 5. PAXOS-CP

In this section, we present an extended version of the Paxos commit protocol. Our extended protocol requires no additional messages to determine the value of an individual log position, and it enables concurrency for read/write transactions that operate on the same transaction group but do not contain conflicting operations. The concurrency is achieved through two enhancements: *combining* concurrent, non-conflicting transactions into a single log position when possible, and when combination is not possible, *promoting* the losing, non-conflicting transactions to compete for the subsequent log position. These enhancement involve changes to Step 3 of the basic Paxos commit protocol.

In Step 3, the Transaction Client examines the votes that have been already cast for the log position. To complete Step 3, the Transaction Client must receive *last vote* responses from a majority of Transaction Services in order to determine the wining value. Let $D$ be the total number of data centers, and let $M = (\lfloor D/2 \rfloor + 1)$ be the minimum number of votes needed to achieve a majority. In the case that the Transaction Client receives exactly $M$ responses, the Transaction Client does not know the votes of the remaining $D - M$ Transaction Services, and so the only safe action is to assume that all of the missing votes are for the same value. If a single value has received a majority of votes, then it is possible that some Transaction Client completed Step 5 of the protocol with this value, and that one or more Transaction Services have written that value into their logs. Therefore, to ensure correctness, the Transaction Client must choose the winning value to be the one with the maximum proposal number (see [18]).

In practice, when a Transaction Client sends a PREPARE message, it will receive responses from more than a simple majority of data centers. In Paxos-CP, the Transaction Client counts the number of votes for each value, and it uses this response information to determine whether to combine, promote, or continue with the basic Paxos commit protocol. The pseudocde for the enhanced protocol is nearly identical to that of the basic protocol. The only change is to replace the call to function `findWinningVal` in Algorithm 2, line 43 with a call to `enhancedFindWinningVal` (lines 76- 87 in Algorithm 2). We explain the combination and promotion enhancements in more detail below.

**Combination.** Let $maxVotes$ be the maximum number of votes for a single value, and let $|responseSet|$ be the number of responses received. The maximum number of votes that any value for a previous proposal could have received is $maxVotes + (D - |responseSet|)$. If this number is less than $M$, then no value has yet received a majority of votes. Therefore, the Transaction Client is free to choose any value for the proposed value in the ACCEPT message.

Instead of simply using its own value, in Paxos-CP, the Transaction Client selects the value to be an ordered list of transactions. To construct this list, the client first adds its own transaction. It then tries adding every subset of transactions from the received votes, in every order, to find the maximum length list of proposed transactions that is one-copy serializable, i.e., no transaction in the list reads a value written by any preceding transaction in the list. With this enhancement, several transactions can be written to the same log position without violating one-copy serializability. While this operation requires a combinatorial number of comparisons, in practice, the number of transactions to compare is small, only two or three. If the number of proposed transactions is large, a simple greedy approach can be used, making one pass over the transaction list and adding each compatible transaction to the winning value.

**Promotion.** The combination enhancement takes advantage of the window in which it is guaranteed that no value has a majority of votes. In the promotion enhancement, the client takes advantage of the situation when a single value *has* received a majority of votes. Again let $maxVotes$ be the maximum number of votes for a single value for log position $k$. If $maxVotes \geq M$, the Paxos protocol guarantees that the value will be written in log position $k$. Therefore, there is no benefit for the client to continue competing for the log position unless its transaction is already part of the winning value. If client's value is not part of the winning value for log position $k$, it can try to win log position $k + 1$ so long as doing so will not violate one-copy serializability. If the client's transaction does not read any value that was written by the winning transactions for log position $k$, the client begins Step 1 of the commit protocol for log position $k + 1$ with its own value. Otherwise, the client stops executing the commit protocol and returns an ABORT status to the application. If the client does not win log position $k+1$, it can try again for promotion to the next log position if one-copy serializability is not violated. As the number of tries increases, there is an increased possibility that the transaction will be aborted because it conflicts with a committed transaction. In practice, this occurs after a small number of promotion attempts, as we show in the evaluations in Section 6.

If the client promotes its transaction to the next log position, or if it detects that it must abort, it stops executing the Paxos protocol before sending ACCEPT messages for the winning value. This early termination does not prevent the winning value from eventually being written to every log. All steps of the Paxos protocol for log position $k$ will eventually be completed, either by another client or by a Transaction Service when it needs to serve a read request.

We now show that Paxos-CP guarantees the necessary properties defined in Section 3 and thus ensures one-copy serializability.

**Theorem 3** *The transactional datastore with replication at multiple data centers, implemented using the transaction protocol and extended Paxos-CP commit protocol defined above, guarantees one-copy serializability.*

PROOF. The transaction protocol remains unchanged, so (A1) and (A2) still hold. The replication guarantee of the Paxos algorithm is not affected by the promotion and combination enhancements. Therefore (R1) is satisfied. The Paxos-CP algorithm also ensures that only committed transactions appear in the log and that all operations from a single transaction appear in a single log position, so properties (L1) and (L2) are also satisfied. The difference between the basic Paxos commit protocol and Paxos-CP relates only to (L3), the requirement that adding an entry to the log preserves one-copy serializability of the history contained in that log. We now show that the combination and promotion enhancements maintain (L3).

Let $H$ be the history in the log up through log position $k-1$, and let $S$ be the serial history that is one-copy equivalent to $H$. Let $t_1, \ldots, t_m$ be the list of transactions that are written to the single log position $k$. Just as in the basic Paxos commit protocol, creating a log entry for any single transaction $t_i$ in the list guarantees one-copy serializability of the log with equivalent serial history $S + t_i$. The combination enhancement ensures that the list of transactions itself is a one-copy serializable history; i.e., for each transaction $t_i$ in the list, $t_i$ does not read from any $t_j$ with $j < i$. This transaction list is one-copy equivalent to the serial history where transactions are ordered as they appear in this list. Therefore, the log with entries up through $k$ is one-copy equivalent to the serial history $S + t_1 + \ldots + t_m$, and (L3) is guaranteed.

In the promotion enhancement, again let $S$ be the one-copy equivalent serial history for the log up through position $k-1$. Let $t_p$ be a transaction that first tries to commit to log position $k$, but is beaten by the transaction (or list of transactions) $t_{w_1}$ and promoted to try for log position $k+1$. It then loses log position $k+1$ to the transaction (or list of transactions) $t_{w_2}$ and is promoted to try for log position $k+2$. This process repeats until $t_p$ eventually commits to log position $k+h$, for some $h > 0$. Let $t_w$ be the list of winning transactions (including $t_p$) for log position $k+h$.

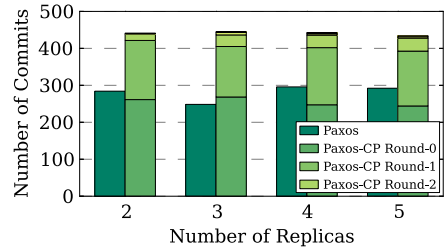

(a) Number of successful transaction commits, out of 500 transactions.

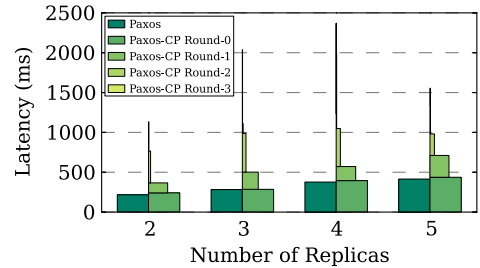

(b) Latency for committed transactions.

**Figure 4: Transaction commits and latency for different numbers of replicas.**

The one-copy equivalent serial history to the log up through position $k+h-1$ is $S + t_{w_1} + \ldots + t_{w_{h-1}}$. In order for $t_p$ to be promoted, it must not have read from any $t_{w_1}, \ldots, t_{w_{h_1}}$, and in order for $t_p$ to be included in the list $t_w$ the list itself must preserve one-copy serializability. Therefore, the log up through position $k+h$ is one-copy equivalent to the serial history $S + t_{w_1} + \ldots + t_{w_{h-1}} + t_w$. This proves that promotion maintains property (L3). □

## 6. EVALUATION

In this section, we present evaluation results of our prototype implementation of the transactional datastore with multi-datacenter replication. Our aim is to compare the performance of the two commit protocols, the basic Paxos commit protocol and Paxos-CP. Therefore, our focus is not on the pure scalability of the system, but rather on the level of transaction concurrency that each protocol allows and the performance trade-offs between the two designs.

Our prototype uses HBase [16] as the key-value store. We have implemented the Transaction Client and Transaction Service (including the commit protocols) in Java. The service is a multi-threaded Java server. UDP is used for communication between the client and remote Transaction Services. We utilize a two second timeout for message loss detection. For communication between a client and its local Transaction Service, we use the optimization described in reference [2]; the client executes HBase operations directly on its local key-value store.

We evaluate our system using the Yahoo! cloud serving benchmark (YCSB) [9]. The benchmark was initially designed for key-value stores and provides no support for transactions. Therefore, we use an extended version of the framework that supports transactions and provides libraries for generating various kinds of transactional workloads [12]. We have implemented the Application Platform as a simple

Java interface between the cloud benchmark application and our Transaction Client.

As our evaluation is focused on the transaction tier, we have simplified the key-value store tier by running a single instance of HBase for each datacenter. Since there is no transactional relationship between different entity groups, we evaluate the transaction protocols on a single entity group consisting of a single row that is replicated in the key-value store in each datacenter. The attribute names and values are generated randomly by the benchmarking framework. Each experiment consists of 500 transactions. Transaction operations are 50% reads and 50% writes, and the attribute for each operation is chosen uniformly at random. We have performed each experiment several times with similar results, and we present the average here.

All evaluations were performed on Amazon's public cloud using medium Hi-CPU instances (*c1.medium*) with Elastic Block Storage. Experiments use between two and five nodes, with each node in a distinct location, or datacenter. Three nodes are in the Virginia region (in distinct availability zones), one node is in Oregon, and one node is in northern California. A single letter for a node indicates the region: **V,O,C**. Round trip time between nodes in Virginia and Oregon or California takes approximately 90 milliseconds. Inter-region communication, Virginia to Virginia, is significantly faster at approximately 1.5 millisecond for a round trip. Round trip time between California and Oregon is about 20 milliseconds.

For Paxos-CP, the combination enhancement has little effect on the performance of Paxos-CP. At most, 24 combinations were performed per experiment, and the average number of combinations was only 6.8 per experiment. We therefore omit a detailed analysis on combinations for space considerations. Transactions were allowed to try for promotion an unlimited number of times. However, as shown in the results below, no transaction was able to execute more than seven promotions before aborting due to a conflict. The majority of transactions commit or abort within two promotions.

**Number of Replicas.** First, we evaluate the performance of the two protocols in systems with different numbers of replica sites. For each run of the experiment, each transaction accesses ten attributes uniformly at random from a row with 100 total attributes. The results, by replica count, are shown in Figure 4. The results for replication in different combinations of datacenters are shown in Figure 5.

In Figure 4(a), we show the commit success count for basic Paxos and each promotion round in Paxos-CP. For the basic Paxos protocol, the mean number of successful transaction commits ranges from 284 out of 500 for the system with two replicas to 292 out of 500 for the system with five replicas. In Paxos-CP, we also see a consistent number of mean total commits (between 434 and 445 out of 500 transactions) regardless of the number of replicas, indicating that replica count has little effect on the commit success rate. We note that, for Paxos-CP, the number of transactions committed in the first round is less than the total number of commits for the basic protocol. This result shows that the promoted transactions are winning out over some first round transactions. When we consider the total number of commits for each protocol, it is evident that the promotion feature leads to a significant increase in the number of successful commits.

Figure 4(b) shows the commit latency for Paxos and each

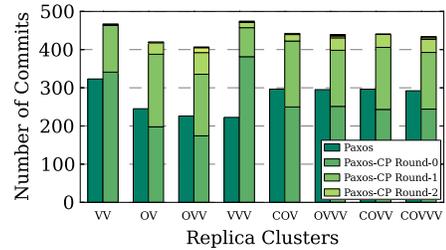

(a) Number of successful transaction commits, out of 500 transactions.

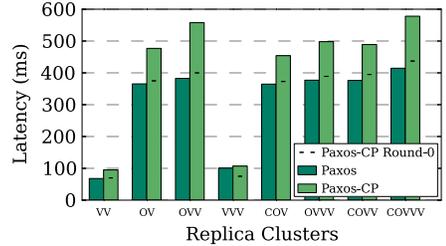

(b) Average latency for all transactions.

**Figure 5: Transaction commits and average transaction latency for different clusters.**

promotion round in Paxos-CP. The first round of Paxos-CP has comparable latency to basic Paxos, with promotions incurring higher latency due the additional rounds of communication required. The latency for each additional promotion round is shown by a stack of narrowing blocks. The decrease in column width is relative to the percentage decrease in the number of commits compared to the previous round. While later rounds experience higher latency, only a small percentage of transactions experience this latency. If increased latency is a concern, the number of promotion attempts can be capped. Both basic Paxos and Paxos-CP exhibit an increase in average transaction latency as the number of replicas increases. While the number of message rounds required to decide a log position does not depend on the number of replicas, a larger number of replicas means more messages per round. There is a increased chance of message loss or delay when the client must contact five replicas (in parallel) instead of two. We believe this contributes to the increased latency observed in the experiments. We note that the transaction latency of the promotion enhancement is lower than would be required for the application to retry an aborted transaction in the basic protocol since a retry would require round trip communication to the datastore to reread the data items in addition to the commit protocol messages.

Figure 5 shows the same experiment as above, broken down by different combinations of datacenters. Figure 5(a) shows the number of commits by promotion round, and Figure 5(b) shows average commit latency for basic Paxos and all rounds of Paxos-CP. Average latency for Paxos-CP transactions completing with no promotions is designated by a small dash. In transactions that involved only Virgina datacenters (VV or VVV) latency is significantly lower, while the improvement on the number of commits for Paxos-CP remains relatively constant despite an inherent increased latency due to location (VV vs. OV) and the lack of a quorum within the same region (VVV vs. COV).

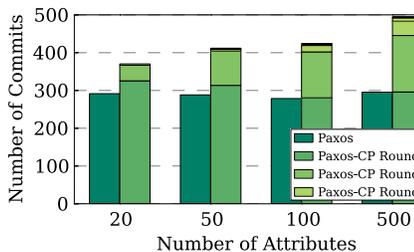
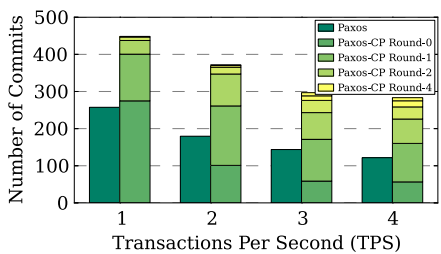
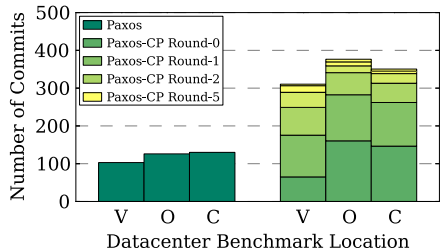

Figure 6: Varying total number of attributes, for three replicas.

Figure 7: Impact of increasing concurrency, for three replicas.

Figure 8: Increasing datacenter concurrency.

**Data Contention.** Next, we investigate the effects of data contention between concurrent transactions for three replica sites. The workload is performed by four concurrent threads with staggered starts, with a target of one transaction per second. Every transaction consists of ten operations, each of which is a read or write on a random attribute in the entity group. We vary the level of contention for individual data items by varying the total number of attributes in the entity group. When the total number of attributes is 20, each transaction accesses 50% of the data items, thus yielding a high level of contention. When the total number of attributes is 500, each transaction accesses only 2% of the data items, resulting in minimal data contention.

The results of this experiment are shown in Figure 6. We use three nodes in the Virginia region with a single YCSB instance. In the basic protocol, no concurrent transaction access is allowed to an entity group regardless of the attributes that are accessed in that transaction. Therefore, the number of transactions that commit successfully is not significantly affected by the level of data contention. For basic Paxos, an average of 290 out of 500 transactions are committed in the worst case (20 total attributes) and 295 out of 500 transactions are committed in the best case (500 total attributes). In contrast, Paxos-CP allows transactions that do not conflict multiple chances to commit, resulting in a higher commit rate. The number of transactions that commit on the first try is similar to the results of the basic protocol. On subsequent attempts, more commit attempts are successful, as shown in Figure 6. The total number of successful commits depends on the data contention. 494 out of 500 transactions committed successfully when data contention was minimal (500 total attributes). Even in the case of high contention (20 total attributes), 370 out of 500 transactions committed, which is 27.5% more than the best case of the basic protocol.

**Increased Concurrency.** Finally, we evaluate the impact of concurrency on the commit success rate. As we increase the number of processes that attempt to update an entity group concurrently, we expect the number of commits to decrease for both protocols due to increased competition for log positions. In Figure 7, we show the effect of increasing the throughput for a single YCSB instance on a VVV replica cluster with contention on 100 attributes. Paxos-CP consistently outperforms basic Paxos in terms of total commits, though both protocols experience a decrease in commits as throughput increases. As throughput increases, promotions play a larger role in Paxos-CP; the increased competition for each log position means that more transactions will be promoted to try for subsequent log positions. We also examine concurrency effects in an experiment where each replica has its own YCSB instance, executing transactions against a shared entity group. In Figure 8, we operate three replicas (VOC). Each YCSB instance attempts 500 transactions over a 100 attribute entity group at a target rate of one transaction per second. Since O and C are geographically closer, a quorum is achieved more easily for these two nodes, resulting in a slightly higher commit rate for their YCSB instances. However, for all datacenters, Paxos-CP has at least a 200% improvement in commits over basic Paxos, while incurring an increase in average latency of 100% for all rounds and 50% increase for the first round latency.

## 7. RELATED WORK

As discussed in this work, Megastore provides both replication across multiple data centers and support for ACID transactions [2]. Windows Azure also provides support for transactions with multi-datacenter replication [6]. It performs replication using a master-slave approach. In the original work [6], replication was not performed on transaction commit; rather, data was replicated every few seconds. Therefore, the system did not guarantee consistency across datacenters. A newer work describes Cloud SQL Server [3], which provides transaction support and consistent replication on top of Windows Azure.

Other recent works have focussed on either transactional support or multi-data center replication, but not both. G-Store implements multi-item transactions where transaction groups are created on demand [10]. Similar to our work, G-Store is implemented on top of a key-value store, but it does not provide replication at multiple data centers.

Google's BigTable [8] provides single item transactions within a single data center. BigTable relies on the files system GFS [14] for data replication. While this replication provides durability, there is still a problem with availability when a BigTable node goes down.

In a recent work, Rao, Shekita and Tata describe Spinnaker [22], a datastore with support for single item transactions and replication within a single data center. Spinnaker also uses Paxos for transaction management and replication. Unlike our system which uses a single instance of Paxos for each log position, Spinnaker uses a Paxos algorithm with a master replica. A leader is elected using Zookeeper [17], and the leader is responsible for ordering all transactions and sending the log entries to all replicas. A leader failure results in a new election. It is not straightforward to extend this approach to support multi-item transactions.

We note that while a leader is alive, the full Paxos algorithm behaves exactly as an atomic broadcast algorithm with a sequencer [13]. One could envision using such a design to implement multi-item transactions with replication using either the full Paxos algorithm or an atomic broadcast protocol like Zab [23]. The leader could act as the transaction manager, check each new transaction against previously committed transactions (in a course grained or fine grained manner) to determine if the transaction can be committed. The leader could then assign the transaction a position in the log and send this log entry to all replicas. Such a design would require fewer rounds of messaging per transaction than in our proposed system, but a greater amount of work would fall on a single site and could possibly be a performance bottleneck. Exploring the tradeoffs between our design and a leader-based approach is a subject for future work.

## 8. CONCLUSION

We have presented a framework for a transactional datastore where data is replicated at multiple datacenters, and we have defined and analyzed the correctness of two transaction management protocols based on the Paxos algorithm. We have shown that the basic Paxos commit protocol acts as a concurrency prevention mechanism rather than a concurrency control mechanism. The enhanced protocol, Paxos-CP provides true transaction concurrency and requires the same per instance message complexity as the original Paxos protocol. Our experiments demonstrate that Paxos-CP achieves a marked improvement in the commit rate over the basic Paxos protocol. As future work, we plan to explore cloud transaction management that uses an optimized Paxos algorithm with a long-term leader.

## 9. ACKNOWLEDGMENTS

This work was funded in part by NSF Grant 1053594, the Arlene & Arnold Goldstein Center at the Technion Autonomous Systems Program, a Technion fellowship, and an Andrew and Erna Finci Viterbi Postdoctoral Fellowship.